\documentstyle[twoside]{article}

\catcode`\@=11
\long\def\@makefntext#1{
\protect\noindent \hbox to 3.2pt {\hskip-.9pt  
$^{{\eightrm\@thefnmark}}$\hfil}#1\hfill}		

\def\thefootnote{\fnsymbol{footnote}}
\def\@makefnmark{\hbox to 0pt{$^{\@thefnmark}$\hss}}	
	
\def\ps@myheadings{\let\@mkboth\@gobbletwo
\def\@oddhead{\hbox{}
\rightmark\hfil\eightrm\thepage}   
\def\@oddfoot{}\def\@evenhead{\eightrm\thepage\hfil
\leftmark\hbox{}}\def\@evenfoot{}
\def\sectionmark##1{}\def\subsectionmark##1{}}

\oddsidemargin=\evensidemargin
\renewcommand{\thefootnote}{\fnsymbol{footnote}}

\newcounter{sectionc}\newcounter{subsectionc}\newcounter{subsubsectionc}
\renewcommand{\section}[1] {\vspace{12pt}\addtocounter{sectionc}{1} 
\setcounter{subsectionc}{0}\setcounter{subsubsectionc}{0}\noindent 
	{\tenbf\thesectionc. #1}\par\vspace{5pt}}
\renewcommand{\subsection}[1] {\vspace{12pt}\addtocounter{subsectionc}{1} 
	\setcounter{subsubsectionc}{0}\noindent 
	{\bf\thesectionc.\thesubsectionc. {\kern1pt \bfit #1}}\par\vspace{5pt}}
\renewcommand{\subsubsection}[1] {\vspace{12pt}\addtocounter{subsubsectionc}{1}
	\noindent{\tenrm\thesectionc.\thesubsectionc.\thesubsubsectionc.
	{\kern1pt \tenit #1}}\par\vspace{5pt}}
\newcommand{\nonumsection}[1] {\vspace{12pt}\noindent{\tenbf #1}
	\par\vspace{5pt}}

\newcounter{appendixc}
\newcounter{subappendixc}[appendixc]
\newcounter{subsubappendixc}[subappendixc]
\renewcommand{\thesubappendixc}{\Alph{appendixc}.\arabic{subappendixc}}
\renewcommand{\thesubsubappendixc}
	{\Alph{appendixc}.\arabic{subappendixc}.\arabic{subsubappendixc}}

\renewcommand{\appendix}[1] {\vspace{12pt}
        \refstepcounter{appendixc}
        \setcounter{figure}{0}
        \setcounter{table}{0}
        \setcounter{lemma}{0}
        \setcounter{theorem}{0}
        \setcounter{corollary}{0}
        \setcounter{definition}{0}
        \setcounter{equation}{0}
        \renewcommand{\thefigure}{\Alph{appendixc}.\arabic{figure}}
        \renewcommand{\thetable}{\Alph{appendixc}.\arabic{table}}
        \renewcommand{\theappendixc}{\Alph{appendixc}}
        \renewcommand{\thelemma}{\Alph{appendixc}.\arabic{lemma}}
        \renewcommand{\thetheorem}{\Alph{appendixc}.\arabic{theorem}}
        \renewcommand{\thedefinition}{\Alph{appendixc}.\arabic{definition}}
        \renewcommand{\thecorollary}{\Alph{appendixc}.\arabic{corollary}}
        \renewcommand{\theequation}{\Alph{appendixc}.\arabic{equation}}
        \noindent{\tenbf Appendix \theappendixc #1}\par\vspace{5pt}}
\newcommand{\subappendix}[1] {\vspace{12pt}
        \refstepcounter{subappendixc}
        \noindent{\bf Appendix \thesubappendixc. {\kern1pt \bfit #1}}
	\par\vspace{5pt}}
\newcommand{\subsubappendix}[1] {\vspace{12pt}
        \refstepcounter{subsubappendixc}
        \noindent{\rm Appendix \thesubsubappendixc. {\kern1pt \tenit #1}}
	\par\vspace{5pt}}

\newcommand{\textlineskip}{\baselineskip=13pt}
\newcommand{\smalllineskip}{\baselineskip=10pt}

\def\eightcirc{
\begin{picture}(0,0)
\put(4.4,1.8){\circle{6.5}}
\end{picture}}
\def\eightcopyright{\eightcirc\kern2.7pt\hbox{\eightrm c}} 

\newcommand{\copyrightheading}[1]
	{\vspace*{-2.5cm}\smalllineskip{\flushleft
	{\footnotesize International Journal of Modern Physics A, #1}\\
	{\footnotesize $\eightcopyright$\, World Scientific Publishing
	 Company}\\
	 }}

\newcommand{\publisher}[2]{{\begin{center}\footnotesize\smalllineskip 
	Received #1\\
	Revised #2
	\end{center}
	}}

\def\abstracts#1#2#3{{
	\centering{\begin{minipage}{4.5in}\baselineskip=10pt\footnotesize
	\parindent=0pt #1\par 
	\parindent=15pt #2\par
	\parindent=15pt #3
	\end{minipage}}\par}} 

\renewenvironment{thebibliography}[1]
	{\frenchspacing
	 \ninerm\baselineskip=11pt
	 \begin{list}{\arabic{enumi}.}
	{\usecounter{enumi}\setlength{\parsep}{0pt}
	 \setlength{\leftmargin 12.7pt}{\rightmargin 0pt} 
	 \setlength{\itemsep}{0pt} \settowidth
	{\labelwidth}{#1.}\sloppy}}{\end{list}}

\newcounter{itemlistc}
\newcounter{romanlistc}
\newcounter{alphlistc}
\newcounter{arabiclistc}

\newcommand{\fcaption}[1]{
        \refstepcounter{figure}
        \setbox\@tempboxa = \hbox{\footnotesize Fig.~\thefigure. #1}
        \ifdim \wd\@tempboxa > 5in
           {\begin{center}
        \parbox{5in}{\footnotesize\smalllineskip Fig.~\thefigure. #1}
            \end{center}}
        \else
             {\begin{center}
             {\footnotesize Fig.~\thefigure. #1}
              \end{center}}
        \fi}

\newcommand{\tcaption}[1]{
        \refstepcounter{table}
        \setbox\@tempboxa = \hbox{\footnotesize Table~\thetable. #1}
        \ifdim \wd\@tempboxa > 5in
           {\begin{center}
        \parbox{5in}{\footnotesize\smalllineskip Table~\thetable. #1}
            \end{center}}
        \else
             {\begin{center}
             {\footnotesize Table~\thetable. #1}
              \end{center}}
        \fi}

\def\@citex[#1]#2{\if@filesw\immediate\write\@auxout
	{\string\citation{#2}}\fi
\def\@citea{}\@cite{\@for\@citeb:=#2\do
	{\@citea\def\@citea{,}\@ifundefined
	{b@\@citeb}{{\bf ?}\@warning
	{Citation `\@citeb' on page \thepage \space undefined}}
	{\csname b@\@citeb\endcsname}}}{#1}}

\newif\if@cghi
\def\cite{\@cghitrue\@ifnextchar [{\@tempswatrue
	\@citex}{\@tempswafalse\@citex[]}}
\def\citelow{\@cghifalse\@ifnextchar [{\@tempswatrue
	\@citex}{\@tempswafalse\@citex[]}}
\def\@cite#1#2{{$\null^{#1}$\if@tempswa\typeout
	{IJCGA warning: optional citation argument 
	ignored: `#2'} \fi}}

\def\pmb#1{\setbox0=\hbox{#1}
	\kern-.025em\copy0\kern-\wd0
	\kern.05em\copy0\kern-\wd0
	\kern-.025em\raise.0433em\box0}

\def\fnt#1#2{\footnotetext{\kern-.3em
	{$^{\mbox{\scriptsize #1}}$}{#2}}}

\def\fpage#1{\begingroup
\voffset=.3in
\thispagestyle{empty}\begin{table}[b]\centerline{\footnotesize #1}
	\end{table}\endgroup}

\def\runninghead#1#2{\pagestyle{myheadings}
\markboth{{\protect\footnotesize\it{\quad #1}}\hfill}
{\hfill{\protect\footnotesize\it{#2\quad}}}}
\font\tenrm=cmr10
\font\tenit=cmti10 
\font\tenbf=cmbx10
\font\bfit=cmbxti10 at 10pt
\font\ninerm=cmr9

\font\eightrm=cmr8

\def\qed{\hbox{${\vcenter{\vbox{			
   \hrule height 0.4pt\hbox{\vrule width 0.4pt height 6pt
   \kern5pt\vrule width 0.4pt}\hrule height 0.4pt}}}$}}

\renewcommand{\thefootnote}{\fnsymbol{footnote}}	

\begin{document}

\runninghead{ Functional Bosonization of Non-Relativistic Fermions
$\ldots$}{ Functional Bosonization of Non-Relativistic Fermions
$\ldots$}
\normalsize\textlineskip
\thispagestyle{empty}
\setcounter{page}{1}

\copyrightheading{}	

\vspace*{0.88truein}

\fpage{1}
\centerline{\bf FUNCTIONAL  BOSONIZATION OF NON-RELATIVISTIC FERMIONS}
\vspace*{0.035truein}
\centerline{\bf IN $(2+1)$ DIMENSIONS}
\vspace*{0.37truein}
\centerline{\footnotesize DANIEL G. BARCI\footnote{e-mail address: 
barci@physics.uiuc.edu or
barci@dft.if.uerj.br}}
\vspace*{0.015truein}
\centerline{\footnotesize\it Department of Physics, University of Illinois at 
Urbana-Champaign, 1110 W. Green Street }
\baselineskip=10pt
\centerline{\footnotesize\it Urbana, Illinois 61801-3080, USA} 
\vspace*{10pt}
\centerline{\footnotesize\it Instituto de F\'\i sica, Universidade do Estado do Rio de Janeiro, 
Rua\ S\~{a}o Francisco Xavier 524}
\baselineskip=10pt
\centerline{\footnotesize\it 20550-013 Rio de Janeiro, RJ, 
Brazil\footnote{Permanent Address}}
\vspace*{10pt}
\centerline{\footnotesize C.\ A.\ LINHARES\footnote{linhares@dft.if.uerj.br}}
\vspace*{0.015truein}
\centerline{\footnotesize\it Instituto de F\'\i sica, Universidade do Estado do Rio de Janeiro, 
Rua\ S\~{a}o Francisco Xavier 524}
\baselineskip=10pt
\centerline{\footnotesize\it 20550-013 Rio de Janeiro, RJ, 
Brazil}
\vspace*{10pt}
\centerline{\footnotesize  J.\ F.\
MEDEIROS NETO\footnote{medeiros@cbpf.br}}
\vspace*{0.015truein}
\centerline{\footnotesize\it Centro Brasileiro de Pesquisas F\'\i sicas, 
Rua Xavier Sigaud 150}
\baselineskip=10pt
\centerline{\footnotesize\it 22290-180 Rio de Janeiro, RJ, Brazil}
\vspace*{10pt}
\centerline{\footnotesize 
A.\ F.\ DE QUEIROZ\footnote{queiroz@dft.if.uerj.br}}
\vspace*{0.015truein}
\centerline{\footnotesize\it Instituto de F\'\i sica, Universidade do Estado do Rio de Janeiro, 
Rua\ S\~{a}o Francisco Xavier 524}
\baselineskip=10pt
\centerline{\footnotesize\it 20550-013 Rio de Janeiro, RJ, 
Brazil}
\vspace*{0.225truein}

\publisher{(received date)}{(revised date)}

\vspace*{0.21truein}
\abstracts{We analyze the universality of the bosonization rules in 
non-relativistic
fermionic systems in $(2+1)d$. We show that, in the case of linear 
fermionic
dispersion relations, a general fermionic theory can be mapped 
into a gauge
theory in such a way that the fermionic density maps into a magnetic flux
and the fermionic current maps into a transverse electric field. These are
universal rules in the sense that they remain valid whatever the interaction
considered. We also show that these rules are universal in the case of
non-linear dispersion relations provided we consider only density--density
interactions. We apply the functional bosonization formalism to a non-relativistic and 
non-local massive 
Thirring-like model and evaluate the spectrum of collective excitations in
several limits. In the large mass limit, we are able to exactly calculate
this spectrum for arbitrary density--density and current--current
interactions. We also analyze the massless case and show that it has no
collective excitations for any density--density potential in the Gaussian
approximation. Moreover, the presence of current interactions may induce a
gapless mode with a linear dispersion relation. 
}{}{}


\vspace*{1pt}\textlineskip	

\section{Introduction}

\vspace*{-0.5pt}
\noindent

\textheight=7.8truein
\setcounter{footnote}{0}
\renewcommand{\thefootnote}{\alph{footnote}}

The method of bosonization has proven to be a very useful technique in
understanding the non-perturbative regime of 
strongly correlated fermionic systems in one spatial dimension. 
Its analysis has shown that the
bosonization rules are universal, in the sense that they are model
independent. Another important property is that the technique displays a
powerful duality mechanism, which is evident in the connection of the
non-perturbative (resp. perturbative) regime of the $(1+1)$-dimensional massive
Thirring model to the perturbative (resp. non-perturbative) one of the
corresponding bosonized sine-Gordon model \cite {Col-Man}.

In recent years,  many efforts have been done in order to
generalize bosonization to higher dimensions, with special emphasis on
$(2+1)d$ fermionic systems, due to their relevance for low-dimensional
condensed-matter physics as, for instance, the quantum Hall effect and
high-$T_c$ superconductivity.

In the context of condensed-matter systems the first attemps to understand
the bosonization of fluctuations around the Fermi surface in higher
dimensions was made by A.\ Luther \cite{Luther} followed by F.\ D.\ M.\ Haldane 
\cite{Haldane},
who considered the two-dimensional system as a collection of infinite one-%
dimensional systems parametrized by an angle. Also, a functional bosonization
method has been applied, essentially, to the evaluation of single-particle
Green's functions of interacting fermionic systems
\cite{Kopietz1}. The fermionic interaction is
decoupled by making a Hubbard--Stratonovich transformation using a bosonic
auxiliary field $\phi(x)$. The Green's function is calculated in a fixed
$\phi(x)$ background and then is averaged using an effective action that
depends only on the Hubbard--Stratonovich field. This approach is suitable
to implement several approximation schemes\cite{Kopietz2} and also
to systematically analyze the case of fermionic systems with non-linear
dispersion relations\cite{Kopietz3}. Bosonization of Fermi liquids
in higher dimensions has been also considered by   A.\ H.\ Castro Neto and 
E.\ Fradkin \cite{Fradkin-Castro-Neto1,Fradkin-Castro-Neto2}. 

As for quantum field theory, two different approaches have been
proposed to bosonize systems in higher dimensions, namely, the canonical and
the functional methods. The canonical method was the one employed to obtain
the first results for the bosonization of free massless fermions in two
dimensions \cite{Marino}. In this case, bosonization is achieved by
introducing a vector gauge field, for which the bosonized action takes the
form of a non-local Maxwell--Chern--Simons model. 

Functional bosonization, as a calculational tool, is undoubtedly a clear
contribution to the understanding of strongly correlated electrons in higher
dimensions. However, it is known in $(1+1)d$ systems that there are general
properties that do not explicitly depend on the details of the one-particle
Green's function, but do depend on global features such as symmetries and
the universality of the fermion--boson mapping. For instance, it has been
shown recently \cite{oxmanmuchiolo} that the conductance of finite
incommensurate Peierls--Fr\"ohlich systems is universal due to a global
chiral anomaly and to the universality of the $(1+1)d$ bosonization rules. For
this reason, it is relevant to ask oneself whether there exist, in higher
dimensions, similar universal rules. 

A first step in this direction has been taken by E.\ Fradkin and F.\ A.\
Schaposnik \cite{FF}, who have analyzed the bosonization of a massive
Thirring model in $2+1$ dimensions. They have shown that in the
infinite-mass limit the fermionic current maps into a topological current,
and the bosonized action is a local Maxwell--Chern--Simons one. Since then,
a great deal of work has been done towards the understanding of the
bosonization structure in $2+1$ dimensions. While the canonical method is
useful to investigate the mapping among the fundamental fields of the
fermionic and the bosonic models, the functional method
\cite{FF,Fidel,BQ1,c1,Banerjee} is appropriate to establish the general
framework for the higher-dimensional bosonization of current correlation
functions.

In a recent paper \cite{bos}, it has been shown for relativistic systems
that the mapping between a fermionic current and a topological bosonic one
is, in fact, universal. This means that this mapping may be performed for
any kind of current interaction. In the present article we extend these
results to the case of non-relativistic interacting fermions. We show that
this universality remains valid in the case of non-relativistic fermions with 
a linearized
dispersion relation. In the case of a non-linear dispersion relation, the
boson--fermion mapping is universal, provided we only consider
density--density interactions. As an example, we apply this technique to the
case of a non-relativistic Thirring-like model with arbitrary density and 
current
interactions. We compute the bosonized action in the Gaussian approximation
and evaluate the spectrum of collective excitations of the model for
arbitrary potentials. We find that in the large mass limit the system has
collective excitations whose dispersion relation has a gap depending on the
particular relation between the potentials. In the masless case, we show
that the model has no collective excitations when considering only density
interactions. However, a gapless collective mode appears when
current--current interactions are taken into account.

The present article in organized as follows. In Section {\bf 2} 
we 
analyze the case of a linearized fermionic
dispersion relation. In \S {\bf 2.1} we define a non-relativistic
Thirring-like model and in \S
 {\bf 2.2} we review the functional
bosonization technique. The Gaussian approximation for the Thirring-like
model is developed in \S {\bf 2.3} and in \S {\bf 2.4} the
two-particle correlation functions are evaluated. The spectrum of collective
excitations for particular limits of the model are described in Section
{\bf 3}. Finally, in Section {\bf 4} we discuss the
universality of the bosonization rules for interacting fermions with a
non-linear dispersion relation and in Section {\bf 5 } we present
our conclusions. 

\section{A non-relativistic Thirring-like model in $(2+1)d$}
\label{linear}


\subsection{The model} 
\label{NLT}

Let us present a non-relativistic (and non-local) Thirring-like model in two
spatial dimensions. The non-relativistic character of the model is displayed
in the kinetic part of the action as well as in the interaction terms. In
order to develop the functional bosonization technique presented in the
following section, we find the Lagrangian formalism more suitable. For this
reason, we present the action of the non-relativistic Thirring-like model
using the imaginary-time ($\tau$) formalism at zero temperature (Euclidean
metric) in the following form\footnote{In what follows we will use the
following notation: Greek indices run from zero to two while Latin indices
may take the values one and two. }:

\begin{equation} 
S =S_0 + S_{\rm int}, \label{free+int} 
\end{equation} 

\noindent with

\begin{equation} 
S_0 = \int_0^\infty d\tau \int d^2x~ \bar{\psi}
\left(\gamma_\mu\partial_\mu+m \right) \psi 
\label{S0} 
\end{equation}

\noindent and
 
\begin{equation} 
S_{\rm int}= \int d\tau_1 d\tau_2 d^2x
d^2y ~ \left[V_{(0)}(x,y) J_0(x) J_0(y) 
+  V(x,y) J_i(x) J_i (y)\right],  
\label{Sint}
\end{equation} 

\noindent where the fermion field $\psi$ is written as 

\[ \psi
=\left(\begin{array}{c} \psi_{1} \\ \psi_{2} \end{array} \right).  
\] 

\noindent The interaction piece of the action has been written in terms
of currents $J_{\mu}$ defined as 

\begin{equation} 
J_{\mu}= \bar{\psi} \gamma_{\mu}\psi 
\left\{ 
\begin{array}{lcl} J_0&=&\psi^{\dagger}\psi \\
 J_i&=&J_i=
\psi^{\dagger}\gamma_0\gamma_i\psi
\end{array}
\right., \label{J} 
\end{equation}

\noindent where we have adopted the representation

\begin{equation}
\gamma_0=-i\sigma_3,~~~\gamma_1=v_1\sigma_1,~~~\gamma_2=v_2\sigma_2,
\end{equation}

\noindent in which $\sigma_1,\sigma_2$ and $\sigma_3$ are the Pauli
matrices, $v_1$ and $v_2$ could be identified with the components of some
``Fermi velocity'', while $V_{0}(x,y)$ and $V(x,y)$ are symmetric bilocal
arbitrary potentials describing density--density interactions and
current--current interactions, respectively. The density--density potential
represents a two-body scattering interaction and may take the form, for
instance, of the usual non-relativistic Coulomb interaction. Although in a
non-relativistic system the current--current potential is much weeker than
the density--density one, we have introduced it here for two reasons. First,
we have considered the potential $V(x,y)$ in order to turn the model more
general, so that we could cover a greater set of situations by choosing,
after the calculations, different potentials (note that in the case of 
$V_0=V=g^2\delta(x-y)$ we recover the usual relativistic local Thirring model). 
Second, in the last few years several models of fermions coupled with 
gauge fields were considered in the description of
strongly correlated electrons \cite{pl}. In particular, the
Chern--Simons dynamics for the gauge field, incorporated in the
Chern--Simons--Landau--Ginzburg theory for the fractional quantum Hall
effect \cite{CSLG}, induces long-distance current--current interactions upon
integration on the transverse gauge fields, which could be modeled by
$V(x,y)$ \cite{eduardo}.

It is not difficult to show that the kinetic term of the action can be
formally manipulated almost in the same way that a relativistc one. In fact,
the algebra of the above defined $\gamma$ matrices, 

\begin{equation}
\left\{\gamma_\mu,\gamma_\nu \right\}=2 g_{\mu\nu},
\end{equation}

\noindent induces a metric tensor $g_{\mu\nu}={\rm diag}(1,v_1,v_2)$. With
this metric, the square of a three-vector $k=(\omega,k_1,k_2)$ reads
$k^2=k_{\mu}k_{\mu}=\omega^2+v_1^2 k_1^2+v_2^2 k_2^2$. Thus, we can work
with a general Euclidean metric $g_{\mu\nu}$ and perform, at the end of the
calculations, an analytic continuation\cite{gelfand} of the metric
coefficients $v_1$ and $v_2$. It is interesting to note that in the
relativistic case of $v_1=v_2=1$ this procedure is nothing but the rigorous
definition of a Wick rotation. Moreover, we shall find interesting
topological results that do not depend on metric, so that without lost of
generality we can work with $v_1=v_2=1$ and make, at the end of the
calculation, the correponding replacements.

The model presented in this section was extensively analyzed in $(1+1)d$ in
the context of the study of many-body physics of quantum wires
\cite{naon-cecilia,naon-virginia,naon-li,bn}. It is the purpose of the present
paper to extend its analysis to $2+1$ dimensions .

\subsection{The bosonization technique} 
\label{bosrel}

In this section we present a functional bosonization technique in $2+1$
dimensions which shall be applied to the particular case of the  model
described in the previous subsection. This technique is suitable for the
calculation of current correlation functions, as it essentially maps a
conserved fermionic current into a topological bosonic one. In $1+1$
dimensions the bosonic field is a scalar one, provided the currents map as 

\begin{equation}
J^F_\mu=\bar\psi \gamma_\mu \psi \longrightarrow
J^B_\mu=\epsilon_{\mu\nu}\partial_\nu\varphi.
\end{equation}

\noindent Note that $J^F_\mu$ is a conserved current ($\partial_\mu
J^F_\mu=0$) due to the global $U(1)$ invariance of the theory; in other
words, it is conserved due to the field equations of motion. On the other
hand, the conservation of the bosonic version of the current ($\partial_\mu
J^B_\mu=0$) is automatic and does not depend on the equations of motion. In
this sense we say that the current is bosonized into a topological one. 

If one wishes to apply the bosonization technique in $2+1$ dimensions in an
analogous fashion, one has to remember the fact that the only way to build
up a topological current in $2+1$ dimensions is by employing a vector field,
instead of a scalar one. Thus, we are looking for a bosonization scheme that
maps fermionic currents into bosonic ones in the following way:

\begin{equation}
J^F_\mu=\bar\psi \gamma_\mu \psi \longrightarrow
J^B_\mu=\epsilon_{\mu\nu\rho}\partial_\nu A_\rho.
\end{equation}

Note that, again in this case, $\partial_\mu J^B_\mu=0$ follows 
automatically, due to
the totally antisymmetric Levi-Civita tensor $\epsilon_{\mu\nu\rho}$ in
$(2+1)d$. It is interesting to remark that the fermionic current is
invariant under a local $U(1)$ gauge transformation of the fields given by

\begin{eqnarray} 
\psi&\longrightarrow& e^{i\alpha(x)}\psi \\ \bar
\psi&\longrightarrow& \bar \psi e^{-i\alpha(x)} ;
\end{eqnarray}

\noindent conversely, the bosonic topological current also has this
invariance, since, when the vector field transforms as

\begin{equation}
A_\mu\longrightarrow A_\mu+i\partial_\mu\alpha(x), 
\end{equation} 

\noindent then

\begin{equation} 
J_\mu'=\epsilon_{\mu\nu\rho}\partial_\nu
(A_\rho+i\partial_\rho\alpha(x))=J_\mu \label{GIC}, 
\end{equation} 

\noindent due to the antisymmetry of $\epsilon_{\mu\nu\rho}$. Therefore, we
expect that, after the bosonization process, the bosonized theory to be a
gauge theory. 

In the functional formalism, the current correlation functions are evaluated
from the generating functional 

\begin{equation}
Z[s]=\int {\cal D}\bar\psi {\cal D}\psi 
e^{-S[\bar\psi,\psi]-i\int d^3x~ s_\mu J^F_\mu} 
\label{ZF}
\end{equation}  

\noindent by functionally differentiating equation (\ref{ZF}) with respect to the
source $s_\mu$: 

\begin{equation} 
\langle J^F_{\mu_1}(x_1)\ldots
J^F_{\mu_n}(x_n)\rangle_S= i^n\frac{\delta\ln Z[s]}{\delta
s_{\mu_n}(x_n)\cdots \delta s_{\mu_1}(x_1)}. \label{correlationF}
\end{equation} 

\noindent The aim of this program is to find a bosonic representation to the
generating functional of current correlation functions in the form 

\begin{equation}
Z[s]=\int {\cal D}A_\mu 
e^{-S_{\rm bos}[A_\mu]-i\int d^3x~ s_\mu J^B_\mu}, 
\label{ZB}
\end{equation}  

\noindent so that 

\begin{equation}
\langle J^B_{\mu_1}(x_1)\ldots J^B_{\mu_n}(x_n)\rangle_{S_{\rm bos}}=
i^n\frac{\delta\ln Z[s]}{\delta s_{\mu_n}(x_n)\cdots \delta s_{\mu_1}(x_1)}.
\label{correlationB}
\end{equation} 

\noindent By comparing eqs. (\ref{correlationF}) and
(\ref{correlationB}) we see that the theories described by $S[\psi]$ and
$S_{\rm bos}[A_\mu]$ are equivalent in the sense that 

\begin{equation}
\langle J^F_{\mu_1}(x_1)\ldots J^F_{\mu_n}(x_n)\rangle_S=
\langle J^B_{\mu_1}(x_1)\ldots J^B_{\mu_n}(x_n)\rangle_{S_{\rm bos}}.
\end{equation}

In order to write a formal expression for the bosonized action and currents,
we follow the path-integral approach of refs. \cite{FF,Fidel,bos}. We
can write the generating functional for interacting massive fermions as 

\begin{equation}
Z[s]=\int {\cal D}\psi {\cal D}
\bar{\psi}~e^{-\int d^3x~\bar{
\psi}(\partial \!\!\!/+m+is\!\!\!/)\psi -S^I[J]},  
\label{intef}
\end{equation}

\noindent where we use the notation $s\!\!\!/=\gamma_\mu s_\mu$ and
$\partial \!\!\!/=\gamma_\mu \partial_\mu$; $S^I[J]$ is an arbitrary
interaction action depending only on fermionic currents. The non-relativistic 
Thirring 
model
presented above is a particular case corresponding to quadratic current
interactions. In what follows we shall consider {\em any } type of current
interactions (not necessarily quadratic). The integration of the fermionic
variables is a very difficult task due to the interaction part of the action
which may contain non-quadratic fermionic terms. The usual trick in the case
of local quartic interactions is to perform a Hubbard--Stratonovich
transformation in order to obtain a quadratic fermionic action, at the
cost of introducing a new auxiliary field. However, for an arbitrary
interaction, such a transformation is not available. Here, we shall obtain a
quadratic fermionic action for arbitrary interacting potentials, provided
that they only depend on the fermionic current. The technique that we will
use was presented for the first time in ref. \cite{bos}. 

We begin by reexpressing the exponential of the fermionic interaction
$S^I[J]$ in terms of its functional Fourier transform, 

\begin{equation}
e^{-S^I[J]}=\int {\cal D}a_\mu e^{-S[a]-i\int d^3x~J_\mu a^\mu }.
\label{FFT}
\end{equation}

\noindent Therefore, for the generating functional of eq.(\ref{intef}),
we get 

\begin{equation}
Z^{\rm int}[s]=\int {\cal D}a_\mu {\cal D}\psi {\cal D}\bar{\psi}~e^{-\int d^3x~
\bar{\psi}(\partial \!\!\!/+m+is\!\!\!/+ia\!\!\!/)\psi -S[a]}.  \label{x}
\end{equation}

\noindent In this way, we can interpret the fermionic interaction as an
effective interacting fermionic system minimally coupled to a dynamical
quantum vector field $a_\mu $ whose action is $S[a]$. Note that for a
generic interaction $S^I[J]$ which is not quadratic in the current
$J_\mu $ the computation of $S[a]$ is, in general, not possible.
However, we will see that our final result will be given only in terms
of the known quantity $S^I[J]$, without relying on the explicit
computation of its Fourier transform $S[a]$. In other words, all that
will be needed is the reasonable assumption that the Fourier
representation (\ref {FFT}) does in fact exist. In particular, for a
local density--density two-body interaction, $S[a]$ is quadratic and
(\ref {FFT}) coincides with the usual Hubbard--Stratonovich
transformation.

The next step towards bosonization is to decouple the external source
$s_\mu$ from the fermion fields. This is accomplished by first making
the following change of variables (with trivial Jacobian)

\begin{eqnarray}
\psi  &\longrightarrow &e^{i\alpha (x)}\psi,   \nonumber \\
\bar{\psi} &\longrightarrow &\bar{\psi}e^{-i\alpha (x)}, \\
{\cal D}\psi{\cal D}\bar\psi &\longrightarrow&
{\cal D}\psi{\cal D}\bar\psi.
\label{jac}
\end{eqnarray}

\noindent The resulting generating functional $Z[s]$ does not depend on
$\alpha$; therefore, we can integrate over $\alpha$ modifying only a
global multiplicative factor that will not enter the correlation
functions. We thus have 

\begin{equation}
Z[s]=\int {\cal D}\psi {\cal D}\bar{\psi}{\cal D}a_\mu {\cal D}\alpha 
\times~ e^{-\int
d^3x~\bar{\psi}(\partial \!\!\!/+i(s\!\!\!/+a\!\!\!/+
\partial \!\!\!/\alpha )+m)\psi-S[a]}.  
\label{p-i}
\end{equation}

\noindent Let us now  introduce 
the field $b_\mu $ through the condition 

\begin{equation}
\partial _\mu \alpha =b_\mu.  
\label{alpha}
\end{equation}

\noindent In order to integrate over $b_\mu $ (instead of $\alpha$) in
the path-integral (\ref{p-i}) we must impose that $b_\mu $ be a
pure-gauge field. This can be done by inserting the delta functional
$\delta (\varepsilon _{\mu \nu \rho }f_{\nu \rho }[b])$ in the
expression for $Z[s]$, $f_{\nu\rho}=\partial_\nu b_\rho- \partial_\rho
b_\nu$ being the $b$ field strength. Thus, 

\begin{equation}
Z[s]=\int {\cal D}\psi {\cal D}\bar{\psi}{\cal D}a_\mu{\cal D}b_\mu ~
\delta
(\varepsilon _{\mu \nu \rho }f_{\nu \rho }[b])~ e^{-\int d^3x~\bar{\psi}
(\partial \!\!\!/+i(s\!\!\!/+a\!\!\!/+b\!\!\!/)+m)\psi-S[a] }.
\end{equation}

The source $s_\mu $ now decouples from the fermions by shifting the
field through $b_\mu $ 

\begin{equation}
b_\mu \longrightarrow b_\mu -s_\mu-a_\mu,  \label{shift}
\end{equation}

\noindent  so that we obtain

\begin{equation}
Z[s]=\int {\cal D}\psi {\cal D}\bar{\psi}{\cal D}a_\mu{\cal D}b_\mu 
~\delta (\varepsilon _{\mu
\nu \rho }f_{\nu \rho }[b-s-a]) ~ e^{-\int d^3x~\bar{\psi}
(\partial \!\!\!/+i b\!\!\!/+m)\psi-S[a] }.
\end{equation}

\noindent The fermionic variables now may be integrated out since their
integral is quadratic and is decoupled from the source $s_\mu$. The
resulting fermionic determinant may be written as an exponential factor,
through the property $\ln \det \hat O= {\rm Tr}\ln \hat O$, and we get

\begin{equation}
Z[s]=\int {\cal D}a_\mu{\cal D}b_\mu ~\delta (\varepsilon _{\mu
\nu \rho }f_{\nu \rho }[b-s-a])~ e^{{\rm Tr}\ln(\partial \!\!\!/+ib\!\!\!/+m)
-S[a]}.
\end{equation}

\noindent Exponentiating the delta functional by means of a Lagrange
multiplier $A_\mu $, the generating functional becomes

\begin{equation}
Z[s]=\int {\cal D}a_\mu {\cal D}A_\mu ~e^{-K_B[A]-i\int
d^3x~\varepsilon _{\mu \nu \rho }A_\mu \partial _\nu (s_\rho +a_\rho
)-S[a]},
\end{equation}

\noindent  where

\begin{equation}
e^{-K_B[A]}=\int {\cal D}b_\mu ~
e^{{\rm Tr}\ln (\partial \!\!\!/+ib\!\!\!/+m)+i\int
d^3x~\varepsilon _{\mu \nu \rho }A_\mu \partial _\nu b_\rho }
\label{KB}
\end{equation}

\noindent isolates the $b$-dependent part of the generating functional.
We now see that the exponential of the bosonizing free action $K_B$ is
obtained from the transverse Fourier transform of the exponential of the
effective action ${\rm Tr}\ln (\partial \!\!\!/+ib\!\!\!/+m)$.

Integrating now over $a_\mu $ and using again the representation 
(\ref{FFT}), we finally  obtain 

\begin{equation}
Z[s]=\int {\cal D}A_\mu ~e^ { -S_{\rm bos}[A_\mu]-i\int d^3x~
s_\mu \varepsilon _{\mu \nu \rho }\partial _\nu A_\rho },  
\label{bosonic}
\end{equation}

\noindent where
 
\begin{equation}
S_{\rm bos}[A_\mu]=K_B[A]+S^I[\varepsilon \partial A]. 
\label{Sbos}
\end{equation}

\noindent As anticipated, we see that our final result relies only on
the known action $S^I$, the unknown action $S[a]$ being only needed at
the intermediate steps. We also see that the bosonized theory is a gauge
theory. This statement is simple to verify. The interaction part of the
action depends only on the current $J_\mu$ that is automatically gauge
invariant as shown in (\ref{GIC}). The gauge invariance of the kinetic
term $K_B$ can be understood from equation (\ref{KB}). If we make a
gauge transformation $A_\mu\rightarrow A_\mu+\partial_\mu \alpha$, the
last term in the exponential of (\ref{KB}) transforms as 

\begin{equation}
\int
d^3x~\varepsilon _{\mu \nu \rho }
\left(A_\mu+\partial_\mu \alpha\right) \partial _\nu b_\rho=\int d^3x~\varepsilon _{\mu \nu \rho }
A_\mu\partial _\nu b_\rho-\int d^3x~\alpha
\varepsilon _{\mu \nu \rho }
 \partial_\mu\partial _\nu b_\rho
\end{equation} 

\noindent where we have integrated by parts the last term. Of course,
this term vanishes, due to the antisymmetry of $\varepsilon_{\mu\nu\rho}$,
from which results the gauge invariance of $K_B$. 

Comparing eqs.~(\ref{intef}) and (\ref{bosonic}), we can read off
the bosonization rules for the fermionic action as well as for the
fermionic current. We see that the bosonic field has the property of a
gauge field, and that the bosonic current takes the form

\begin{equation} 
J_\mu=\epsilon_{\mu\nu\rho}\partial_\nu A_\rho \left\{
\begin{array}{lcl} J_0&=&B \\ J_i&=&\epsilon_{ij} E_j 
\end{array}
\right. . 
\end{equation} 

\noindent Therefore, the fermionic density bosonizes to a ``magnetic'' field
$B$ and the fermionic current to a transverse ``electric'' field $E_i$,
both related to the $A_\mu$-field. These are the main results of this section
and they are summarized in Table \ref{rules}.

\begin{table}
\[
\begin{array}{|l||c|c|} 
\hline 
         &     \mbox{{\it Fermionic}} & \mbox{{\it Bosonic}} \\
\hline \hline
\mbox{{\it Fundamental Fields}} &   \psi=\left( \begin{array}{c}
						\psi_1 \\
						\psi_2 
						\end{array}\right) &
A_\mu   \\
\hline
\mbox{{\it Conserved Current}} & 
 \psi^{\dagger}\gamma_0 \gamma_\mu \psi  &
\epsilon_{\mu\nu\rho}\partial_\nu A_\rho  \\
\hline 
\mbox{{\it Kinetic Term}} & 
\int d^3x~\bar{
\psi}(\partial \!\!\!/+m)\psi
&   K_B(A_\mu)  \\
\hline
\mbox{{\it Interaction Term}} &
S^I(\bar{\psi}\gamma_\mu \psi )&   S^I(\varepsilon_{\mu \nu
\rho }\partial _\nu A_\rho )\\
\hline
\end{array}
\]
\tcaption{Bosonization rules for fermions with linearized dispersion 
relation in $2+1$ dimensions}
\label{rules}
\end{table}

Another interesting and important result is that the kinetic term and
the interaction term of the fermionic action are bosonized
independently. That is to say, the bosonization process does not mix
these terms. Last but not least, the equivalence between the fermionic
and the bosonic current is not only exact but it is also a solid
universal result, in the sense that it remain the same in the
interacting quantum theory. We have shown that we can bosonize any term
involving the fermionic current by simply replacing $J_{\mu}$ by the
bosonic topological current $\varepsilon _{\mu \nu \rho }\partial _\nu
A_\rho $.

The universality for such a large class of interaction terms shown in this
section is strongly dependent on the fact that the kinetic theory has only
first order derivatives. We shall come back to this point in \S {\bf 4.2}.

\subsection{The Gaussian approximation}
\label{Gaussian}

As we have seen, the results depicted in table \ref{rules} are exact and
universal. In other words, they do not depend on any particular detail
of the fermionic interaction provided  that it depends only on the fermionic
vector bilinear. In this sense, we have exactly bosonized the current
and a general class of interactions. However, the kinetic fermionic
action is bosonized into $K_B[A_\mu]$ given by eq. (\ref{KB}). Thus, in
order to obtain 
an explicit expression for 
this action we have to evaluate the transverse Fourier
transform of the exponential of the fermionic determinant. Of course,
this is a very difficult task and no exact result is known in two
spatial dimensions. Nevertheless, expressions (\ref{KB}),
(\ref{bosonic}) and (\ref{Sbos}) constitute the starting point for
different approximation schemes not always available in the fermionic
version. For example, the semiclassical approximation is very easily
performed in a bosonic theory, while it is very hard to handle in the
fermionic counterpart. 

A very interesting observation about the action $K_B$ was made in ref.
\cite{bos}. It was shown that there is a non-linear and non-local
field transformation that maps the bosonized action $K_{B}$ into a pure
local Chern--Simons form. 
The underlying
topological structure of the bosonized action 
is then made explicit
and, in some sense, this
transformation is a convenient form to distinguish the topological
contents of a fermionic theory from the non-topological dynamical
contributions. However, although this formal observation is a clear
positive contribution towards the understanding of the bosonized action,
the practical applications of such a structure as a reliable computational
tool remain undeveloped. 

Therefore, it is necessary to develop an approximation scheme for the
evaluation of current correlation functions. In ref. \cite{Fidel},
the limit in which the ``mass'' parameter $m$ is taken to infinity 
was considered. This is
equivalent to a low-energy expansion of the fermionic theory, since the
energy fluctuations must be much smaller than
the typical energy of the gap, $2m$. In this limit, the bosonized
kinetic term takes the form of a local Maxwell--Chern--Simons term:

\begin{equation} 
K_B[A_\mu]=\int d^3x
\frac{i}{2}\epsilon_{\mu\nu\rho}A_\mu\partial_\nu A_\rho +\frac{1}{4
m}F_{\mu\nu}F_{\mu\nu} +\cdots. 
\label{KBinfty}
\end{equation} 

\noindent This is a quadratic action, simplifying in this way the
evaluation of current correlation functions. It is worth stressing that in
this limit the bosonized action is exact, in the sense that the next
quartic term is proportional to $1/m^2$. 

In ref. \cite{bof}, a more general approximation was discussed,
the so-called Gaussian approximation, in which one takes into account
the full quadratic part of the fermionic determinant, which may be
written as

\begin{equation}
{\rm Tr}\ln (\partial \!\!\!/+ib\!\!\!/+m) 
\, =\,  T_{PC}(b) \; + \; T_{PV}(b)
\label{Det2}
\end{equation}

\noindent with

\begin{eqnarray}
T_{PC}(b)\,&=& \,-\frac{1}{4} \,\int d^3 x \, 
F_{\mu \nu}(b)\, F(-\partial^2) \; F_{\mu \nu}(b),
\label{PC} \\
T_{PV}(b)\,&=&\,-\frac{i}{2} \,\int d^3 x \,  b_{\mu} \, G(-\partial^2)
\, \epsilon_{\mu \nu \lambda}\partial_{\nu}b_{\lambda},
\label{PV}
\end{eqnarray}

\noindent where $T_{PC}$ and $T_{PV}$ come from the parity-conserving
and parity-violating pieces of the vacuum-polarization tensor,
respectively\cite{loop}.
 
A standard one-loop calculation for the Fourier transforms of the
functions $F$ and $G$ in (\ref{Det2}) yields \cite{bof}

\begin{eqnarray}
{\tilde F} (k^2) &=& \frac{\mid m \mid}{4 \pi k^2} 
\left[ 1 - \displaystyle{\frac{1 \,-\,\displaystyle{\frac{k^2}{4 m^2}}}{(
\displaystyle{\frac{k^2}{4 m^2}})^{\frac{1}{2}}}}  \arcsin(1+
\frac{4 m^2}{k^2})^{-\frac{1}{2}} \right] 
\nonumber  \\
{\tilde G} (k^2)&=& \frac{m}{2 \pi \mid k \mid}
\, \arcsin (1 + \frac{4 m^2}{k^2} )^{- \frac{1}{2}},
\label{FG}
\end{eqnarray}

\noindent where here and in what follows we shall always denote
momentum-space representations by putting a tilde over the corresponding
coordinate-space representation quantity.

In order to evaluate $K_B$, it is necessary to calculate the functional
transverse Fourier transform of (\ref{Det2}). Since this expression is
quadratic, it is a simple task to evaluate the functional integral
provided we have properly fixed the gauge. We thus obtain

\begin{eqnarray}
K_{B}=\int d^3 x
\left\{  \frac{1}{4} F_{\mu \nu} \, C_1 \, F_{\mu \nu}
- \frac{i}{2} A_{\mu} \, C_2 \, \epsilon_{\mu \nu \lambda}
\partial_{\nu} A_{\lambda} \right\},
\label{KB2}
\end{eqnarray}
where
\begin{eqnarray}
C_1 &=&\frac{F}{ - \partial^2 F^2 \, + \, G^2 }, \label{c1}\\
C_2 &=& \frac{G}{ - \partial^2 F^2 \, + \, G^2 }. \label{c2}
\end{eqnarray}

In the limit $m\rightarrow \infty$ we recover expression
(\ref{KBinfty}). Although in this limit the quadratic approximation is
exact, it is not obvious how to determine its range of validity for
small $m$. However it was shown in refs. \cite{bos} and
\cite{foscoex} that this quadratic approximation is in fact the
exact one for any ``mass'', provided we use it to calculate two-point
correlations functions. For higher order correlation functions it is not
at all easy to justify this approximation for small $m$.

Turning now to the interaction term of the model, its bosonization
is performed by just replacing $J_0\rightarrow
\frac{1}{2}\epsilon_{ij}F_{ij}$ and $J_i\rightarrow
\frac{1}{2}\epsilon_{i\mu\nu}F_{\mu\nu}$ in (\ref{Sint}). We thus obtain

\begin{equation}
S^I[A_\mu]=\int d^3x d^3y ~ \left\{\epsilon_{ij}F_{ij}(x) V_{(0)}(x,y)
\epsilon_{lm}F_{lm}(y)
+F_{i0}(x)V_{(1)}(x,y)F_{i0}(y) \right\}.
\label{SIA}
\end{equation}

Finally, from eqs.~(\ref{KB2}) and (\ref{SIA}) we can write the
bosonized action for the non-relativistic Thirring model in $(2+1)$ dimensions
in the Gaussian approximation as

\begin{eqnarray}
S_{\rm bos}[A]&=&\int d^3x~ \left\{\frac{1}{4} F_{\mu \nu} \, C_1 \, F_{\mu \nu}
- \frac{i}{2} A_{\mu} \, C_2 \, \epsilon_{\mu \nu \lambda}
\partial_{\nu} A_{\lambda}\right\} \nonumber \\ 
&+&\int d^3x d^3y ~ \left\{\frac{1}{4}\epsilon_{ij}F_{ij}(x) V_0(x,y)
\epsilon_{lm}F_{lm}(y) + F_{i0}(x)V(x,y)F_{i0}(y)\right\}. 
\label{SB}
\end{eqnarray}

\subsection{Current--density correlation functions}
\label{correlations}

In order to calculate current--current correlation functions,
we need to evaluate the generating functional $Z[s]$ given by
(\ref{bosonic}) and (\ref{Sbos}). We shall use for $S_{\rm bos}$ the
Gaussian approximation obtained in (\ref{SB}). 

Since we are dealing with a gauge theory, to functional integrate the
generating functional we have to properly fix the gauge in (\ref{SB}).
We gain more insight of the problem by choosing the Coulomb gauge
$\vec{\nabla}\cdot\vec{A}=0$ and reexpressing the gauge-fixed action in
terms of $A_0$ and the associated magnetic field $B=\vec\nabla\times\vec
A$.
 
We note that, since $\vec{\nabla}\cdot\vec A=0$, we can always write the
components of the field $\vec A$ as 

\begin{equation}
A_i=\epsilon_{ij}\partial_j \phi,
\label{Aphi}
\end{equation}

\noindent where $\phi(x)$ is an arbitrary scalar field. The relation
between $B$ and $\vec A$ implies, together with (\ref{Aphi}), that $B$
is none other that the Laplacian of the field $\phi$ (apart from a
sign), so that the vector potential is expressed in terms of $B$ using
the Laplacian Green's function:

\begin{equation}
A_i=-\epsilon_{ij}\frac{\partial_j}{\nabla^2}B.
\label{Ai}
\end{equation}

\noindent In this way we can write the generating functional in the
Coulomb gauge as a function of $A_0$ and $B$:

\begin{equation}
Z[s]=\int {\cal D}A_0 {\cal D}B~e^{-S_b(A_0,B)-i\int d^3x~s_0 B+s_i\left(\epsilon_{ij}\partial_j A_0
-\frac{\partial_0\partial_i}{\nabla^2}B\right)},
\end{equation}

\noindent with

\begin{eqnarray}
 S_{\rm bos}&=& 
\frac{1}{2}\int d^3xd^3y\left[ B(x) 
\left\{ C_1
\left(1+\frac{\partial_0^2}{\nabla^2}\right)+
\left(V_0+V   \frac{\partial_0^2}{\nabla^2} \right)\right\}(x-y)
B(y) \right.
\nonumber\\
&-& \left.\frac{1}{2} A_0(x) \left\{\nabla^2
\left(C_1+V\right)\right\}(x-y)  A_0(y)
-i A_0(x) C_2(x-y) B(y)\right],
\label{SA0B}
\end{eqnarray}

\noindent where $C_1$ and $C_2$ are given by (\ref {c1}) and (\ref{c2})
and $V_0(x-y)$ and $V(x-y)$ are the density--density and current--current
interaction potentials, respectively.

At this stage of the calculation, it is convenient to write the action
in momentum space. Thus, by Fourier transforming (\ref{SA0B}), we obtain

\begin{eqnarray}
S_{\rm bos}
&=&\frac{1}{2}\int \frac{d\omega d\vec{k}}{(2\pi)^3}~ 
  \tilde B^* 
\left\{ \tilde C_1\left(1+\frac{\omega^2}{|\vec{k}|^2}\right)+
\left(\tilde V_0+\tilde V   \frac{\omega^2}{|\vec k|^2} \right) 
\right\}\tilde B 
\nonumber\\ 
&&-\frac{1}{2}\int \frac{d\omega d\vec{k}}{(2\pi)^3}~ 
\tilde A_0^* \left\{\vec k^2
\left(\tilde C_1+\tilde V\right)\right\} 
\tilde A_0 \nonumber \\
&&-i \int \frac{d\omega d\vec{k}}{(2\pi)^3}~ 
\tilde A_0(\omega,\vec{k}) \tilde C_2 \tilde B(-\omega,-\vec{k}).
\end{eqnarray}

\noindent Integrating out the $\tilde A_0$ and $\tilde B$ fields, we
find

\begin{equation}
Z[s]=
e^{-\int d\omega d^2k~ \tilde s_\mu(\omega,\vec{k}) \Pi_{\mu\nu}(\omega,\vec k)
\tilde s_\nu(-\omega,-\vec k)},
\label{zeta}
\end{equation}

\noindent where the vacuum polarization tensor $\Pi_{\mu\nu}$ takes the
form

\begin{equation}
\Pi_{\mu,\nu}(\omega,\vec k)={\cal P}(\omega,\vec k)
\left(\delta_{\mu\nu}-\frac{k_\mu k_\nu}{k^2}\right)
+i{\cal Q}(\omega,\vec k)\epsilon_{\mu\nu\rho}k_\rho,
\label{PI}
\end{equation}

\noindent with 

\begin{eqnarray}
{\cal P}(\omega,\vec k)&=& \left(\omega^2+k^2\right)
\left(\tilde V+\tilde C_1\right)\Delta(\omega,k) \\
{\cal Q}(\omega,\vec k)&=& \tilde C_2 \Delta(\omega,k)
\end{eqnarray}

\noindent and

\begin{equation}
\tilde\Delta(\omega,\vec k)= 
\frac{1}
{\left(\tilde V+\tilde C_1 \right)
\left\{ \omega^2\left(\tilde C_1+\tilde V \right)+ 
\vec k^2\left(\tilde C_1+\tilde V_0 \right)\right\}
+\tilde C_2^2}.
\label{Delta}
\end{equation}

The vacuum polarization tensor for the interacting theory $\Pi_{\mu\nu}$
has the correct symmetry properties. In fact, 

\begin{equation}
\Pi_{\mu\nu}(\omega,\vec k) =
\Pi_{\nu\mu}^*(\omega,\vec k)=\Pi_{\nu\mu}(-\omega,-\vec k);
\end{equation}

\noindent also, it must be transverse, as a consequence of
gauge invariance:

\begin{equation}
k_\mu \Pi_{\mu\nu}= \Pi_{\nu\mu}k_\mu=0.
\label{TRANS}
\end{equation}

\noindent This condition is automatically satisfied due to the tensor
structure of $\Pi_{\mu\nu}$ (see eq.~(\ref{PI})).

It is interesting to note that we can rewrite the parity-conserving part
of the vacuum polarization tensor as 

\begin{equation}
{\cal P}(\omega,\vec k)= 
 \frac{\omega^2+k^2}{
\left\{ 
\omega^2\left(1+\frac{\tilde V}{\tilde C_1} \right)+ 
\vec k^2\left(1+\frac{\tilde V_0}{\tilde C_1} \right)
\right\}
+\frac{\tilde C_2^2}{\tilde C_1}
\frac{1}{\left(1+\frac{\tilde V}{\tilde C_1}\right)}
}.
\end{equation}

\noindent This formula should be compared with the one-spatial-dimensional
result \cite{bn}. We see that the effect of the 
kinematics in two space dimensions, given by $\tilde C_1$ and $\tilde C_2$,
is twofold. The last
term in the denominator (proportional to $\tilde C_2$) is due to a
parity-breaking effect and we shall show that it is responsible for
opening gaps in the excitation spectrum (see Section {\bf 3}).
The parity-conserving effects renormalize the potentials as $\tilde
V^R=\frac{\tilde V}{\tilde C_1}$ and $\tilde V_0^R=\frac{\tilde
V}{\tilde C_1}$. Note that this is a dynamical renormalization (since
$\tilde C_1$ depends on $\omega$), which changes the structure of the
interactions. However, we shall see that, in the large $m$ limit  $\tilde
C_1$ is a constant and the net effect is just a coupling-constant
renormalization. 
 
Taking into account this structure for $\Pi_{\mu\nu}$, 
it is now simple to write an
expression for the current expectation value as we have, by definition,

\begin{equation}
\langle  J_\mu\rangle =\frac{\delta \ln Z[s]}{\delta s_\mu};
\end{equation}

\noindent from eq. (\ref{zeta}), it is expressed in terms of
$\Pi_{\mu\nu}$ in the form 

\begin{equation}
\langle  J_\mu\rangle=\Pi_{\mu\nu}s_\nu  
= \Pi_{\mu 0} s_0 + \Pi_{\mu i} s_i.
\end{equation}

\noindent Gauge invariance implies that the components of $\Pi_{\mu\nu}$ are
not independent (see eq.~(\ref{TRANS})). In particular, we have

\begin{equation}
\Pi_{\mu 0}=-\frac{1}{\omega}\Pi_{\mu i} k_i.
\end{equation}

\noindent Using this relation and defining an external electric field as 
$\vec E^{\rm ext}=\vec\nabla s_0-\partial_0 \vec s$, we find 

\begin{equation}
\langle  J_\mu\rangle =i\frac{1}{\omega} \Pi_{\mu j}\tilde E^{\rm ext}_j.
\label{Jota}
\end{equation}

It is interesting to have an explicit equation for the density
fluctuations (in the linear response approximation) induced by an
arbitrary external electromagnetic field. We obtain from (\ref{PI}) and
(\ref{Jota})

\begin{eqnarray}
\langle \rho\rangle &=&i\frac{1}{\omega} \Pi_{0 j}\tilde E^{\rm ext}_j 
 \nonumber\\
&=&- \frac{i}{\omega^2+\vec k^2}{\cal P}(\omega,\vec k)~ k_i 
\tilde E^{\rm ext}_i
+\frac{1}{\omega}{\cal Q}(\omega,\vec k)~ 
\epsilon_{ij} k_i \tilde E^{\rm ext}_j 
\nonumber \\
&=& -\tilde \Delta(\omega,\vec k)(\tilde V+\tilde C_1)
{\cal F}\left(\vec\nabla \cdot \vec E^{\rm ext} \right)-i
\frac{\tilde C_2 \tilde\Delta(\omega,\vec k)}{\omega}{\cal F}
\left(\vec\nabla\times 
\vec E^{\rm ext} \right),
\end{eqnarray}

\noindent where ${\cal F}(\ldots)$ denotes Fourier transform. 

Noting that the {\em external classical field $\vec E^{\rm ext}$}
satisfies Maxwell's equations, we finally obtain 

\begin{equation}
\langle \rho\rangle=-\tilde\Delta(\omega,\vec k)
\left\{\tilde \rho^{\rm ext}(\tilde V+\tilde C_1)+\tilde C_2 
\tilde B^{\rm ext}           
\right\}.
\label{density}
\end{equation}

\noindent We see that a density-charge excitation in
the system 
may arise as a consequence of external charges and/or external magnetic
fields. The general relation between a magnetic field and density comes,
in general, from the parity-breaking terms of the model. In the
fermionic version this is seen as a gap in the spectrum and, in the
bosonic picture, as a Chern--Simons-like action. Equation
(\ref{density}) is a generalization of the Aharonov--Casher calculation
\cite{Aharonov-Casher} to interacting systems submitted to an arbitrary
electromagnetic field. 

Similarly to $\langle\rho\rangle$, we can write a expression for
$\langle J_i\rangle$: 

\begin{equation}
\langle J_i\rangle =i\left\{\frac{{\cal P}(\omega,\vec k)}{\omega}~\tilde 
E_i^{\rm ext} +i 
{\cal Q}(\omega,\vec k)~ \epsilon_{ij} \tilde E_j^{\rm ext}\right\}
-\frac{k_i {\cal P}(\omega,\vec k)}{\omega(\omega^2+\vec k^2)}
~\tilde \rho^{\rm ext}.
\end{equation} 

\noindent In the case of a divergenceless external electric field, we
find

\begin{equation}
\langle J_i(\omega,\vec k)\rangle=
\sigma_{ij}(\omega,\vec k) \tilde E_{j}^{\rm ext}(\omega,\vec k), 
\end{equation}

\noindent where we have introduced a conductivity tensor

\begin{equation}
\sigma_{ij}= \left(\begin{array}{cc}
\sigma_{xx} & \sigma_{xy}  \\
-\sigma_{xy} & \sigma_{yy}
\end{array}  \right),
\end{equation}

\noindent for whose elements we have

\begin{eqnarray}
\sigma_{xx}(\omega,\vec k)&=&\sigma_{yy}(\omega,\vec k)=
\frac{{\cal P}(\omega,\vec k)}{\omega},   \\
\sigma_{xy}(\omega,\vec k)&=& {\cal Q}(\omega,\vec k).
\end{eqnarray}

In the same way we have calculated expressions for the current mean
value, we can write expressions for the two-point current correlation
function, as it takes the simple form 

\begin{equation}
\langle  J_\mu(x) J_\nu(y)\rangle =
\frac{\delta \ln Z[s]}{\delta s(y)\delta s(x)}
=\Pi_{\mu\nu}(x-y).
\end{equation}

\noindent In particular, using (\ref{PI}), the density--density
correlation function may be written as 

\begin{equation}
\langle \rho\rho \rangle = \Pi_{00}=\vec k^2 (\tilde V+\tilde C_1) 
\tilde\Delta(\omega,\vec k).
\label{rho-rho}
\end{equation}

\noindent The excitation spectrum is given by the analytic properties of
$\tilde\Delta$. In the next section we shall develop this point in
detail.

Similarly, the spatial components of the two-point current correlation
function also take a simple form: 

\begin{eqnarray}
\langle J_1 J_1 \rangle &=& (\tilde V+\tilde C_1)(\omega^2+ k_2^2) 
\tilde \Delta(\omega,\vec k), \label{J1J1} \\
\langle J_2 J_2 \rangle &=& (\tilde V+\tilde C_1)(\omega^2+ k_1^2) 
\tilde \Delta(\omega,\vec k), \label{J2J2} \\
\langle J_1 J_2 \rangle &=& -\left\{(\tilde V+\tilde C_1)k_1k_2 
+\tilde C_2 \omega\right\}\tilde \Delta(\omega,\vec k). \label{J1J2}
\end{eqnarray}

\section{Spectrum of collective excitations}
\label{spectrum}

\subsection{The large $m$ limit}

We have seen from the density--density correlation function (eq.
(\ref{rho-rho})) that the spectrum of collective excitations can be read
off from the singularities of $\tilde\Delta(\omega,\vec k)$ in the
complex $\omega$ plane. Let us analyze here the case of a large
fermionic gap where the simplest case to be considered is the
infinite-gap limit. Performing the $m\rightarrow\infty$ limit in 
eq.~(\ref{FG}), we obtain

\begin{eqnarray}
\lim_{m\rightarrow \infty} \tilde F(k)&=& 0, \\
\lim_{m\rightarrow \infty} \tilde G(k)&=&\frac{1}{4\pi},
\end{eqnarray}

\noindent implying  $\tilde C_1=0$ and $\tilde C_2=(1/2\pi)^3$.
Replacing these values into eq. (\ref{Delta}), we obtain

\begin{equation}
\tilde \Delta(\omega,\vec k)= \frac{1}{\tilde V^2}
\frac{1}{\omega^2 +\vec k^2 \frac{\tilde V_0}{\tilde V}+
\frac{2}{(2\pi)^6\tilde V^2}}.
\label{del1}
\end{equation}

\noindent In the infinite-gap limit in the absence of current--current
interactions, $\tilde\Delta$ does not depend on $\omega$, which implies
the nonexistence of any kind of density excitation. Moreover, the
conductivity of the system is $\sigma_{xx}=\sigma_{yy}=0$, while
$\sigma_{xy}=1/2$ (in units of $e^2/h$). This result is completely
independent of the density--density potential $V_0$. The presence of a
transverse current comes from the fact that the fermionic system is not
invariant under a parity transformation due to the term proportional to
$m\bar\psi\psi$. As a consequence, the system behaves as an insulator
($\sigma_{xx}=0$) whatever the potential $V_0$; although there is a
transverse current, it does not dissipate energy since it is
perpendicular to the applied electric field. It is important to stress
that this is the same behavior of free fermions with a large gap, as
the effective action for the external electromagnetic field is a pure
topological Chern--Simons one. Through this simple exercise, we thus
verify that the same system but with an arbitrary two-body interaction
($\rho$--$\rho$) remains topological in the infinite-gap limit. 

Turning back to the case in which the current--current interaction is
present, still in the infinite gap limit, we realize that it plays a
crucial role. The potential $V$ induces a pole in the density--density
correlation function associated to a collective excitation with the
dispersion relation

\begin{equation}
\omega=\pm \sqrt{\frac{\tilde V_0}{\tilde V}\vec k^2+
\frac{\tilde C_2^2}{\tilde V^2}}
\end{equation}

\noindent and residue 

\begin{equation}
{\rm Res}(\tilde \Delta)=
\frac{1}{2\sqrt{\tilde V_0\tilde V\vec k^2+\tilde C_2}}.
\end{equation}

\noindent Consider, for example, the case of a local current--current
interaction with coupling constant $g^2$ and a Coulomb density--density
interaction $\tilde V_0=e^2/|k|$. In this case, the collective excitations
have a gap proportional to $1/g^2$, so that, the stronger the current
interactions, the weaker the energy necessary to excite the collective
modes. A direct consequence of the existence of this gap is that the
conductivity is not affected by these excitations. It is simple to verify
that also with a strong current interaction, $\sigma_{xx}=0$ and
$\sigma_{xy}=1/2$. This topological property breaks down if the gap
disappears as it happens with a potential $\tilde V_0=1/|k|^\alpha$ with
$\alpha\ge 2$. 

A different situation emerges when the fermionic gap $2m$ is
large but not infinity. We now keep a $1/m$-contribution from eq. (\ref{FG}), 
so that $\tilde \Delta$ is given by eq.~(\ref{Delta}) with
$\tilde C_1= \frac{1}{12\times16 \pi^2 m}$ and $\tilde C_2=(1/2\pi)^3$.
In these conditions, the espectrum of collective excitations contains a
mode with dispersion relation

\begin{equation}
\omega=\pm \sqrt{\frac{\tilde V_0+\tilde C_1}{\tilde V+\tilde C_1}\vec k^2+
\frac{\tilde C_2^2}{(\tilde V+\tilde C_1)^2}      }
\label{disp1}
\end{equation}

\noindent and residue

\begin{equation}
{\rm Res}(\tilde \Delta)= \frac{1}{2
\sqrt{(\tilde V+\tilde C_1)(\tilde V_0+\tilde C_1)\vec k^2+
\frac{\tilde C_2^2}{(\tilde V+\tilde C_1)}}}.
\end{equation}

\noindent As expected, this case has a similar behavior to the previous
one ($m\rightarrow\infty$) with the only difference that the potentials
$\tilde V_0$ and $\tilde V_1$ are renormalized by the constant $\tilde
C_1$. In fact, the parity-conserving contribution of the dynamics of the
free fermions acts as local density--density and current--current
interactions with coupling constant $\tilde C_1$ renormalizing in this
way the ``true'' potentials. As a final remark, note that these results
are exact for any interaction. In particular, eq. (\ref{disp1})
represents the exact dispersion relation of the collective modes for any
arbitrary static potentials $\tilde V_0(\vec k)$ and $\tilde V(\vec k)$.

\subsection{The massless case}

In this section we analyse the $m\rightarrow 0$ limit of the non-relativistic Thirring
model. Making  $m\rightarrow 0$ in (\ref{FG}) we find

\begin{eqnarray}
\lim_{m\rightarrow 0}\tilde F(\omega,\vec k)
&=&\frac{1}{16\sqrt{\vec k^2-\omega^2}},     \\
\lim_{m\rightarrow 0}\tilde G(\omega,\vec k)&=&0,
\label{Gm=0}
\end{eqnarray} 

\noindent implying that
 
\begin{eqnarray}
\lim_{m\rightarrow 0}\tilde C_1(\omega,\vec k)
&=&\frac{16}{\sqrt{\vec k^2-\omega^2}},     \\
\lim_{m\rightarrow 0}\tilde C_2(\omega,\vec k)&=&0.
\label{C2m=0}
\end{eqnarray}

\noindent This result leads to 

\begin{equation}
\tilde \Delta=\frac{1}{16\sqrt{\vec k^2-\omega^2}-
\tilde V \omega+\vec k^2 \tilde V_0}.
\label{Deltam=0}
\end{equation}

Here, several comments are in order. First of all, expression
(\ref{Deltam=0}) depends essentially on eq.~(\ref{Gm=0}), which is not
well established and it seems to be regularization dependent \cite{reg}.
In ref.\ \cite{bos}, we argue that if we use a regularization that
does not introduce any additional symmetry breaking on the system, the
correct result is that of eq.~(\ref{Gm=0}). Fortunately, although the
value of $\tilde\Delta$ depends on the particular value of $\lim_{m\to
0}\tilde G$, it was shown in ref. \cite{asym} that the
collective-excitation spectrum does not depend on it, being insensible
to the parity-breaking term of the fermionic determinant.

To understand the excitation spectrum, let us simplify expression
(\ref{Deltam=0}) by considering a pure density--density interaction
($\tilde V=0$). In this case, eq.~(\ref{Deltam=0}) reduces to 

\begin{equation}
\tilde \Delta=\frac{1}{16\sqrt{\vec k^2-\omega^2}+\vec k^2 \tilde V_0}.
\label{Deltam=0V=0}
\end{equation}

\noindent This equation for $\tilde \Delta$ has no poles. Instead, it
has a cut for $\vec k^2< \omega^2$ whose 
discontinuity 

\begin{equation}
\delta^+\tilde\Delta\equiv-i\lim_{\epsilon\to 0}
\left\{\tilde\Delta(\omega+i\epsilon,\vec k)-
\tilde\Delta(\omega-i\epsilon,\vec k) \right\}
\label{deltaDelta}
\end{equation}

\noindent is given by

\begin{equation}
\delta^+\tilde\Delta=
\frac{16\sqrt{\omega^2-\vec k^2}}{16^2(\omega^2-\vec k^2)+
\vec k^4 \tilde V_0^2}\Theta(\omega^2-\vec k^2)\Theta(\omega),
\end{equation} 

\noindent where $\Theta$ is the Heaviside function. This result shows
that the model in the zero mass limit (in the Gaussian
approximation) has no collective excitations. Instead, there is a
continuum of modes that contribute to any physical observable. For
instance, from eq.~(\ref{density}) it is possible to calculate the
density fluctuations induced by an external charge. However, this
density profile cannot be interpreted as a superposition of density
waves with a definite dispersion relation \cite{NLW}.

Let us consider now the effect of a current--current interaction on this
system. For simplicity, let us consider $\tilde V_0=0$ and $\tilde
V=g^2$ in (\ref{Deltam=0}). We immediately realize that we have two
regimes of momenta. For $\vec{k}^2<\omega^2$ we have a cut in the
spectrum whose discontinuity is given by 

\begin{equation}
\delta^+\tilde\Delta=
\frac{16\sqrt{\omega^2-\vec k^2}}{16^2(\omega^2-\vec k^2)+
g^4 \omega^2}\Theta(\omega^2-\vec k^2)\Theta(\omega).
\end{equation} 

\noindent Moreover, for high momenta $\vec{k}^2>\omega^2$, the current-
current potential induces a gapless mode with linear dispersion relation
$\omega=v|\vec k|$ with $v=\frac{1}{1+g^2/16^2}< 1$.

\section{Non-relativistic spinless electrons in $(2+1)d$}
\label{nonlinear}
\subsection{The current generating functional}

In this section we consider a spinless non-relativistic
fermionic system in two spatial dimensions 
with kinetic action given by 

\begin{equation}
S_0=\int d^2x dt~\psi^*(x)
\left\{i\partial_t+\frac{1}{2m}\nabla^2-\mu\right\}\psi(x).
\end{equation} 

\noindent This action has a global $U(1)$ symmetry and as a consequence
it has conserved Noether currents,

\begin{eqnarray}
J_0&=&\psi^*\psi,      \label{J0}  \\
J_i&=&\psi^*\nabla_i \psi, \label{Ji}
\end{eqnarray} 

\noindent in such a way that $\partial_t J_0+\vec\nabla\cdot\vec J=0$.

In principle, any functional of the currents $S_I(J_0,J_i)$, could be
considered for the interaction part of the action, thus we can write
quite generally

\begin{equation}
S[\psi^*,\psi]=S_0[\psi^*,\psi]+S_I[J_0,J_i] 
\label{action}
\end{equation}

\noindent The idea is to develop the bosonization technique discussed is
the last section to evaluate density-current correlation functions. The
main point of this technique is to couple the system to an external
electromagnetic field and use it as a source to evaluate current
expectation values. In the present case, this is not so simple to
accomplish, due to the non-linear dispersion relation ($\omega=\vec
k^2/2m$). More explicitly, let us couple the action (\ref{action}) to a
vector gauge field $(s_0,\vec s)$. The result is

\begin{equation}
S=\int d^2x dt~\psi^*
\left\{i\partial_t+s_0+\frac{(\vec\nabla+\vec s)^2}{2m}-
\mu\right\}\psi 
+ S_I(J_0,\vec J+\vec s J_0).
\label{actions}
\end{equation} 

\noindent The main difference with respect to the linear case is that
here the interaction action is modified due to the derivative operator
that appears in the current. In other words, $J_i$ is not a local 
gauge-invariant current and therefore it picks up a factor $s_i$ which couples
to $J_0$. By functional differentiating (\ref{actions}) with relation to
$(s_0,s_i)$, we find 

\begin{eqnarray}
\frac{\delta S}{\delta s_0}&=&J_0,   \label{source1}\\
\frac{\delta S}{\delta s_i}&=&J_i+\frac{J_0}{2m}s_i +
\frac{\delta S_I}{\delta J_i} J_0.  
\label{source2}
\end{eqnarray}

\noindent We see from (\ref{source1}) that in the limit
$s_\mu\rightarrow 0$ the potential $s_0$ may be interpreted as a source
of $J_0$. However, from (\ref{source2}) it can be deduced that in the
same limit $s_i$ only acts as a source of $J_i$ provided 

\begin{equation}
\frac{\delta S_I[J_0,J_i]}{\delta J_i}=0,
\end{equation} 

\noindent implying that, in order to use the functional bosonization
technique developed in the present article for the evaluation of current
correlation functions, we must consider only density--density
interactions.

\subsection{The bosonization technique}
\label{nonrel}

Consider the generating functional
\begin{equation}
Z[s]=\int {\cal D}\psi{\cal D}\psi^*~e^{-S[\psi^*,\psi,s_\mu]},
\end{equation}

\noindent where

\begin{equation}
S[\psi^*,\psi,s_\mu]=\int d^2x dt~\psi^*
\left\{i\partial_t+s_0+\frac{(\vec\nabla+\vec s)^2}{2m}-
\mu\right\}\psi + S_I[J_0].
\end{equation}

\noindent In order to integrate the fermions, we first rewrite the
interaction action as 

\begin{equation}
e^{-S^I[J_0]}=\int {\cal D}\phi 
e^{-\tilde S[\phi]-i\int d^3x~J_0 \phi},
\label{phi}
\end{equation}

\noindent so that 

\begin{equation}
Z[s]=\int {\cal D}\psi{\cal D}\psi^*{\cal D}\phi~
e^{-S[\psi^*,\psi,s_\mu,\phi]}
\end{equation}

\noindent where now we have

\begin{equation}
S=\int d^2x dt\psi^*
\left\{i\partial_t+s_0+i\phi+\frac{(\vec\nabla+\vec s)^2}{2m}-
\mu\right\}\psi+\tilde S[\phi].
\end{equation}

\noindent The decoupling of the source $s_\mu$ from the fermions and
expressing the generating functional as a function of a gauge field
$A_\mu$ follow from the same formal mathematical steps that one takes
from eq. (\ref{jac}) to eq.~(\ref{Sbos}), obtaining

\begin{equation}
Z[s]=\int {\cal D}A_\mu ~e^ { -S_{\rm bos}[A_\mu]-i\int d^3x~
s_0  B +s_i\varepsilon_{ij}E_j },  
\label{bosonicnr}
\end{equation}

\noindent where 

\begin{equation}
S_{\rm bos}[A_\mu]=K_B[A]+S_I[B]
\label{Sbosnonrel}
\end{equation}

\noindent and

\begin{equation}
e^{-K_B[A]}=\int {\cal D}b_\mu ~
e^{{\rm Tr}\ln\left(i\partial_t+b_0+\frac{(\vec\nabla+\vec b)^2}{2m}-
\mu\right)+i\int
d^3x\varepsilon _{\mu \nu \rho }A_\mu \partial _\nu b_\rho }.
\label{KBnonrel}
\end{equation}

In this way, for a non-relativistic gas with density--density interactions,
we find bosonization rules very similar to the ones encountered for the
fermionic system with linear dispersion relation. In fact, the exponential
of the bosonized free fermion action is the transverse Fourier transform of
the fermionic determinant and the bosonized interaction action has the same
form as the fermionic interacting action when substituting the density
$\rho$ by the magnetic field $B$. Moreover, from the coupling to an external
source $s_\mu$, it is simple to read off the bosonization rules for the
currents: 

\begin{equation}
J_\mu=\epsilon_{\mu\nu\rho}\partial_\nu A_\rho
\left\{
\begin{array}{lcl}
J_0&=&B \\ 
J_i&=&\epsilon_{ij} E_j
\end{array} 
\right. .
\end{equation}

\noindent This means that we can map any fermionic theory in $2d$ into a
gauge theory by associating charges to magnetic fluxes and currents to
transverse electric fields. These results are summarized in Table
\ref{rulesNL}.


\begin{table}
\[
\begin{array}{|l||c|c|} 
\hline 
         &     \mbox{{\it Fermionic}} & \mbox{{\it Bosonic}} \\
\hline \hline
\mbox{{\it Fields}} &   \psi, \psi^*  &
A_\mu   \\
\hline
\mbox{{\it    Density}} & 
 \psi^*\psi  &      B  \\
\hline 
\mbox{{\it   Current}} & i\psi^*\nabla_i\psi & \epsilon_{i,j} E_j
\\
\hline
\mbox{{\it Kinetic Term}} & 
\int dx\psi^*
\{i\partial_t+\frac{\nabla^2}{2m}-\mu\}\psi
&   K_B(A_\mu)  \\
\hline
\mbox{{\it Density Int.}} &
S^I(\rho)&   S^I(B)\\
\hline
\mbox{{\it Current Int.}} &
S^I(J_i)&  \mbox{Not Universal}\\
\hline
\end{array}
\]
\tcaption{Bosonization rules for fermions with quadratic  dispersion 
relation in $2+1$ dimensions}
\label{rulesNL}
\end{table}


Eq.~(\ref{KBnonrel}) for $K_B$ is an exact formal result and can be
used to develop different kind of approximation schemes. As an example,
let us consider here the Gaussian approximation. 

In the Gaussian approximation the fermionic determinant takes the form 

\begin{eqnarray}
\lefteqn{{\rm Tr}\ln\left(i\partial_t+b_0+
\frac{1}{2m}\left(\vec\nabla+\vec b\right)^2-
\mu\right)=} \nonumber \\
&&n\int d^3x~ b_0(x)-
\int d^3x d^3y~ b_\mu(x){\cal K}_{\mu\nu}(x-y)b_\nu(y),
\end{eqnarray}

\noindent where $n$ is the mean density of electrons. The first (linear)
term is a tadpole contribution and can be dropped out simply by
measuring densities from $n$. The second term, the non-relativistic
vacuum polarization tensor, gets the contribution from two types of
diagrams, the bubble and the tadpole diagrams arising form the
interaction $\vec b^2 \psi^*\psi$. It is simple to evaluate $K_B$, since
the transverse Fourier transform is quadratic. Thus, after a proper
gauge fixing is introduced, we can perform the functional integral,
obtaining

\begin{equation}
K_B[A]=\frac{1}{4}\int d^3k~ \tilde B^*
{\cal K}_{00}^{-1}(\omega,\vec k) \tilde B+
\frac{1}{4}\int d^3k~ \tilde E^*_i 
\left(\frac{k_i k_j}{k_ik_j {\cal K}_{ij}-\vec k^2{\rm Tr}{\cal K}}\right)
\tilde E_j.
\end{equation}
 
\noindent We have used the fact that the vacuum polarization tensor is
symmetric in this approximation, which implies that the free system does not
break parity. For this reason, a cross-term $B E_i$ is absent from $K_B$.
However, it may happen that configurations with strong magnetic fields, that
explicitly break parity, could induce a Chern--Simons-type action. In 
refs.\ \cite{sakita} and \cite{luca-barci-aragao}, this behavior
of the functional determinant was deduced for long distances; the resulting
action is non-quadratic, making the evaluation of the transverse Fourier
transform a very difficult task. However, it is possible to perform it in
the limit of a strong magnetic field, in which it is possible to project the system
onto the first Landau level\cite{universal}. 

Turning back to the Gaussian approximation, we note that it has only
been performed in the bosonized action of the free fermionic term. The
interaction part is exactly bosonized by just replacing $\rho$ by $B$.
Considering a two-body scattering interaction $\frac{1}{4}\rho V\rho$,
the full bosonized action reads 

\begin{equation}
S_{\rm bos}[A]=\frac{1}{4}\int d^3k~ \tilde B^*
\left(\frac{1+{\cal K}_{00}\tilde V(\vec k)}{{\cal K}_{00}}\right) 
 \tilde B+
\frac{1}{4}\int d^3k~ \tilde E^*_i 
\left(\frac{k_i k_j}{k_ik_j {\cal K}_{ij}-\vec k^2{\rm Tr}{\cal K}}\right)
\tilde E_j.
\end{equation} 

Note that we were able to transform a fermionic interacting action in an
action that describes dynamical electric fields and magnetic fluxes. In
order to calculate current correlation functions we must fix the gauge.
In the Coulomb gauge, the above action reads

\begin{equation}
S_{\rm bos}[A]=\frac{1}{4}\int d^3k~ \tilde B^*
\left(\frac{1+{\cal K}_{00}\tilde V(\vec k)}{{\cal K}_{00}}\right) 
 \tilde B-
\frac{1}{4}\int d^3k~ \tilde A_0^* 
\left(\frac{|\vec k|^4}{k_i k_j {\cal K}_{ij}-\vec k^2{\rm Tr}{\cal K}}\right)
\tilde A_0.
\end{equation} 
  
It is now a simple task to evaluate, for example, the density--density 
correlation function:

\begin{equation}
\langle\rho(\omega,\vec k) \rho(-\omega,-\vec k) \rangle_{S}=
\langle B(\omega,\vec k) B(-\omega,-\vec k) \rangle_{S_{\rm bos}}=
\frac{{\cal K}_{00}}{1+{\cal K}_{00}\tilde V(\vec k)}.
\label{roro}
\end{equation} 

\noindent It is interesting to note, that, in this case, the Gaussian
approximation coincides with the random-phase approximation (RPA). Of
course, equation (\ref{KBnonrel}) is a natural framework to
systematically improve this approximation. 

The current--current corelation function reads 

\begin{eqnarray}
\lefteqn{\langle J_i(\omega,\vec k) J_k(-\omega,-\vec k)\rangle_{S} }
 \nonumber\\
&=&\epsilon_{il}\epsilon_{km}
\langle E_i(\omega,\vec k) E_k(-\omega,-\vec k)\rangle_{S_{\rm bos}}=
\langle \vec E^2\rangle \delta_{ik}-\langle E_iE_k\rangle \nonumber \\
&=&\left(\vec k^2 \delta_{ik}-k_ik_k \right)
\langle A_0A_0\rangle+\frac{\omega^2}{\vec k^2}\frac{k_ik_k}{\vec k^2}
\langle B B\rangle \nonumber \\
&=&\left(\vec k^2 \delta_{ik}-k_ik_k \right)
\frac{k_lk_n {\cal K}_{ln}-\vec k^2 {\rm Tr} {\cal K}}{\vec k ^4}
+
\frac{\omega^2}{\vec k^2}\frac{k_ik_k}{\vec k^2}
\frac{{\cal K}_{00}}{1+{\cal K}_{00}\tilde V(\vec k)}.
\label{jj} 
\end{eqnarray}

\noindent Finally, we can evaluate the density--current correlation
function, obtaining

\begin{eqnarray}
\langle \rho(\omega,\vec k) J_k(-\omega,-\vec k)\rangle_{S}&=&\epsilon_{km}
\langle B(\omega,\vec k) E_k(-\omega,-\vec k)\rangle_{S_{\rm bos}}=\frac{\omega k_k}{\vec k^2}
\langle B B\rangle \nonumber \\
&=&\frac{\omega k_k}{\vec k^2}
\frac{{\cal K}_{00}}{1+{\cal K}_{00}\tilde V(\vec k)}.
\label{roj}
\end{eqnarray}

For the sake of consistency let us check whether current conservation
$k_i J_i+\omega \rho =0$ holds in the above current correlation
functions. Multiplying eq. (\ref{jj}) by $k_i$ and using eq.
(\ref{roj}), we obtain

\begin{equation}
k_i\langle J_i(\omega,\vec k) J_k(-\omega,-\vec k)\rangle_{S}=
\frac{\omega^2}{\vec k^2}k_k
\frac{{\cal K}_{00}}{1+{\cal K}_{00}\tilde V(\vec k)}
=\omega \langle \rho(\omega,\vec k) J_k(-\omega,-\vec k)\rangle_{S},
\end{equation}

\noindent as it must be. Let us remark that the current conservation in
the fermionic version is guaranteed thanks to the global $U(1)$
invariance, while in the bosonic counterpart it is guaranteed by the
topological character of the current.

\section{Conclusions}
\label{discussions}

In this paper we have addressed 
the question of the universality of current bosonization
rules in $(2+1)d$ non-relativistic fermions with 
linear and non-linear dispersion relations. We have shown that
in the linear dispersion relation case, the fermionic action bosonizes
to a gauge theory. The bosonization process does not mix the kinetic and
interacting parts of the action, making it possible to analyze them
individually. We have found that the kinetic part of the fermionic
action maps into an effective action $K_B[A]$ that is given by the
logarithm of the transverse Fourier transform of the fermionic
determinant. Conversely, the interacting part of the action bosonizes by
simply replacing the fermionic current $J_\mu$ by the topological
bosonic one $J_\mu=\epsilon_{\mu\nu\rho}\partial_\nu A_\rho$. These
rules are universal, in the sense that they do not depend on the
particular model. Moreover, as the bosonized current is topological, it
is metric independent and therefore this  rule does not
depend on any local property of the model.

In the case of a non-linear dispersion relation, we have found that the
universal rules can be applied if we only consider
density--density interactions. The current--current interaction destroys
universality in this case. The reason for this behavior resides in the
fact that for non-linear dispersion relations, the fermionic currents
(for instance $j_i=\psi^*\nabla_i\psi$) are not gauge invariant. For
this reason, we believe that it could be possible to find universal
bosonization rules in the general case, provided we are able to
reformulate the bosonization technique in terms of gauge-invariant
objects. 
 
We have applied the functional bosonization formalism to a non-relativistic
Thirring-like model and have evaluated the spectrum of collective
excitations in various limits. In the large-mass limit, we were able to
exactly calculate this spectrum for arbitrary density--density and
current--current interactions. We have found collective excitations whose
dispersion relation has a gap depending on the particular relation
between the potentials. An interesting observation is that the 
parity-conserving part of the vacuum polarization tensor (and of course, the
expression for the collective excitations) is similar to the 
one-dimensional Tomonaga--Luttinger model, up to a coupling constant
renormalization. The physical interpretation is that the kinetic energy
of  fermions in two spatial dimensional with a large gap (mass $m$) is frozen.
Consequently, the only effect of kinematics is to renormalize the
coupling constants. Of course, in $(2+1)$ dimensions there is, in addition
to the parity-conserving term, a parity-breaking term $C_2$, which has
no equivalent in (1+1) dimension and is the responsible for opening the
gap in the excitation spectrum. It is necessary to stress here that
these are exact results without any kind of approximation other than the
consideration of a large gap responsible for the insulating face. 

We have also analyzed the massless case in a Gaussian approximation in order 
to evaluate the bosonized action and we have shown that in this approximation
there are no collective excitations for any density--density potential.
However, the presence of a current interaction may induce a gapless
collective mode with linear dispersion relation.

Another important application of bosonization is to improve the usual
approximation schemes in many body physics. Since we are bosonizing the
interactions exactly, all the approximations are made on the bosonized
kinetic fermionic action. In order to improve the Gaussian
approximation, we need a better understanding of the structure of the
transverse Fourier transform of the fermionic determinant. A first step
towards this aim was made in refs. \cite{bos} and
\cite{determinante} where a topological Chern--Simons structure
was identified. In this context, the problem of going beyond the
quadratic approximation in $K_B$ is that of evaluating the Fourier
transform of an interacting gauge theory. An interesting method to
decouple this type interaction was recently proposed in ref.
\cite{foscofildel}.
 
Summarizing, we believe that the existence of universal rules for
bosonization in two spatial dimensions could help us improve our
understanding of strongly correlated systems. In particular, the
topological character of the current and the topological content of the
bosonized action could play a similar r\^ole to the one that chiral
symmetry plays in one dimension, possibly allowing, for instance, the
calculation of universal properties of two dimensional systems. This
kind of application is now under development and will be reported
elsewhere \cite{universal}. Of course, a definite improvement on the
comprehension of the structure of $K_B$ would be to discover
a fermion--boson mapping between fields (and not between
currents) in two dimensions. This is one of the most important open
questions within this subject.

\nonumsection{Acknowledgements}
\noindent
We would like to acknowledge Profs. S.\ P.\ Sorella and L.\ E.\ Oxman
for useful discussions. D.\ G.\ B.\  also acknowledges
Prof.\  E.\ Fradkin for useful comments and for his kindly hospitality
at UIUC, where part of this work was completed. 

D.\ G.\ B.\ is partially suported by State University of Rio de Janeiro, 
the Brazilian agency CNPq though a post-doctoral fellowship and  
NSF, grant number 
NSF DMR98-17941 at UIUC. 

The Conselho Nacional de Desenvolvimento Cient\'\i fico e Tecnol\'ogico,
the Funda\c {c}\~{a}o de Amparo \`{a} Pesquisa do Estado do Rio de
Janeiro, and SR2-UERJ are gratefully acknowledged for financial
support.


\end{document}